\documentclass [12pt]{article}
\usepackage {epsfig}

\begin{document}

\begin{center}
{\Large {\bf Seasonal Modulations of the Underground Cosmic-Ray Muon Energy}}

\vskip 0.5cm
{\bf
A.S.~Malgin
}
~~~

{\it
Institute for Nuclear Research RAS, 117312, prospect 
60-letya Oktyabrya,7a, Moscow, Russia
}
\end{center}

\noindent E-mail: Malgin@lngs.infn.it

\begin{abstract}
The parameters of the seasonal modulations (variations) in the intensity of muons 
and cosmogenic neutrons generated by them at a mean muon energy of 
280 GeV have been determined in the LVD (Large Volume Detector) 
experiment. The modulations of muons and neutrons are caused by a 
temperature effect, the seasonal temperature and density variations 
of the upper atmospheric layers. The analysis performed here leads 
to the conclusion that the variations in the mean energy of the muon 
flux are the main source of underground cosmogenic neutron variations, 
because the energy of muons is more sensitive to the temperature 
effect than their intensity. The parameters of the seasonal modulations 
in the mean energy of muons and the flux of cosmogenic neutrons at the 
LVD depth have been determined from the data obtained over seven years 
of LVD operation.
\end{abstract}

\noindent {\bf Keywords:} atmospheric muons, neutron yield, 
                          underground experiment

\vskip 0.5cm
\noindent
{\bf
A full version was published in Journal of Experimental 
and Theoretical Physics, 2015, Vol. 121, No. 2, pp. 212-216.
}
\vskip 1.5cm

\section{Introduction}
\label{intro}

At present, the correlation between the annual modulations in the 
counting rate of events observed in dark matter particle search 
experiments \cite{Ber14}, \cite{Ral10}, \cite{Blu11}, \cite{Dav14} 
and the seasonal variations in the underground muon flux is being 
actively discussed. The muon intensity variations at great depths 
are considered as a possible source of the seasonal modulations of 
events in low-background detectors. It is assumed that the modulations 
can be produced by cosmogenic neutrons whose flux is linearly 
related to the varying muon flux.

The cause of the seasonal muon variations at sea level and underground 
is well known \cite{Bla38}, \cite{Bar52}, \cite{Dor56}. This is a 
temperature effect that leads to a change in the density of the 
terrestrial atmosphere and its height as a result of its heating 
in summer and cooling in winter. On the one hand, a decrease in the 
density of the upper atmospheric layers (stratosphere) through 
expansion when heated leads to an increase in the probability of 
the ${\pi} \to {\mu}$ decays of first generation charged pions 
from extensive air showers (EASs) and to a corresponding decrease 
in the number of pions (and the number of their decays ${\pi} \to {\mu}$) 
in the last generations. On the other hand, the atmosphere expansion 
increases the probability of the ${\mu} \to  e$ decays of low energy 
muons on their way to the Earth. The first fact, an increase in the 
probability of the decays ${\pi} \to {\mu}$, gives a positive temperature 
effect observed in the high energy muon flux. The last two facts 
associated with low energy muons lead to a negative temperature 
effect, a reduction in the muon intensity at sea level, where the mean 
muon energy is $\sim 4$ GeV. The observed muon variations are determined 
by the combined action of the negative and positive effects. The 
negative component dominates approximately down to 20 meters water 
equivalent (m w.e.). Its contribution is decreased with a depth 
increasing and becomes negligible starting from approximately 
200 m w.e. ($E_{\mu} = 35 $ GeV). Muons at energies above 1 TeV 
produced at $pA$-interaction energies above 100 TeV can reach 
depths greater than 2 km w.e. The positive temperature effect is 
clearly observed at large underground facilities \cite{Amb97},
 \cite{Sel09}, \cite{Dan11}.

In the LVD \cite{Aga11} experiment the parameters of seasonal variations 
both muon intensity and number of neutrons produced by muons for a 
fixed time interval (60 days) in LVD structure were determined. 
The puzzle of the data obtained is inequality of amplitudes of 
the neutron number modulation and modulation of muon intensity: 
the first is about 10 times greater than the second one. This 
problem can be solved by assuming seasonal variations of 
the muon flux mean energy which determines the neutron production. 

The positive temperature effect increases the probability of the 
decays of first pion generations in EAS. This must lead not only 
to a rise in the intensity of muons at great depths but also to an 
increase in their mean energy. Below, we will estimate the seasonal 
variations in the mean muon energy and the variations in the number 
of neutrons produced by muons using the LVD data \cite{Sel09}, 
\cite{Aga11}.

\section{The Determining of the number of muon-induced neutrons in the LVD}

The Large Volume Detector (LVD) is described in detail in 
\cite{Sel09}, \cite{Bar88}, \cite{Agl92}. The main objective of the 
detector is to search for neutrino bursts from the gravitational 
stellar core collapse.

The muon intensity at the LVD depth is 
$I^0_{\mu} = (3.31 \pm 0.03) \times 10^{-4} m^{-2} s^{-1}$ \cite{Agl98}. 
The threshold muon energy to reach the LVD depth is 1.3 TeV \cite{Bar88}. 
The ionization losses of a vertical muon in the LVD structure are, on average,
$ \sim 2.2$ GeV. The criteria for the selection of muon events from all the detected 
events specify the muon rate in each tower $\sim 1.2$ min$^{-1}$. The 
mean energies of single muons and muons in pairs are $(270 \pm 18)$ 
and $(381 \pm 21)$ GeV, respectively \cite{Amb03}. Single muons 
account for $ \sim 90 \%$ of the number of muon events. The mean 
energy of the muon flux is $\overline E_{\mu}$ = 280 GeV. Below, we will 
use this value.

Neutrons are produced by muons in the LVD scintillator and the 
elements of its steel structure. The detection efficiency of 
neutrons produced in the scintillator and uniformly distributed 
in a counter volume is $\sim 50\%$. Neutrons produced in iron 
are detected with an efficiency of $\sim 20\%$. The delayed 
coincidence method is used to determine the number of generated 
neutrons: $\gamma$-quanta from capture of a neutron by a 
free proton in the scintillator or by an iron nucleus in the 
steel structure are detected in a time interval of 1 ms after 
the muon passage through the LVD.

The counters through which the muon passed and those adjacent 
to them are included in the analysis. The time distribution 
of pulses summed over all muons and counters is described by 
a function $N(t) =N_0exp(-t/{\tau}_{\gamma}) + B$. $N_0$ is 
determined at constant B specifying the background of measurements 
and known time ${\tau}_{\gamma} = 180 {\mu}s$. The product 
$N_0{\tau}_{\gamma}$ gives the number of detected ${\gamma}$-quanta 
in the time interval from 0 to $\infty$ Using $N_0{\tau}_{\gamma}$
 and taking into account the neutron detection efficiency 
(including the ${\gamma}$-quanta detection efficiency), 
we obtain the number of neutrons produced in the selected muon events.

\section{Seasonal modulations in the mean energy of the muon 
         flux and muon-induced neutrons}

The time dependence $I_{\mu}(t)$ of the muon flux per day over 
eight years of LVD operation ($21.5 \times 10^6$ muon events) 
starting from January 1, 2001 (Fig. 1a) was obtained in \cite{Sel09}:
\begin{equation}
I^{\mu}(t) = I^{\mu}_0 + {\delta}I^{\mu}cos(\frac{2{\pi}}{T}(t-t^{\mu}_0)). 
\end{equation}
The mean intensity was $(3.31 \pm 0.03) \times 10^{-4} m^{-2} s^{-1}$; 
the modulation period was $T = (367 \pm  15)$ days. The phase 
$t^{\mu}_0  = (185 \pm 15)$ days corresponds to a maximum muon 
intensity at the beginning of July. The intensity modulation 
amplitude is ${\delta}I^{\mu} = (5.0 \pm 0.2) \times 10^{-6}m^{-2}s^{-1}$. 
The derived modulation parameters are consistent with the measurements 
in the MACRO experiment at the same depth as the LVD one \cite{Amb97}. 
The muon intensity measurements at the BOREXINO facility 
(2007 - 2011, $4.6 \times 10^6$ muons) located near the LVD 
\cite{Dan11} are also consistent with the LVD and MACRO parameters.

 \begin{figure}[!t]
  \centering
  \includegraphics[width=4.0in]{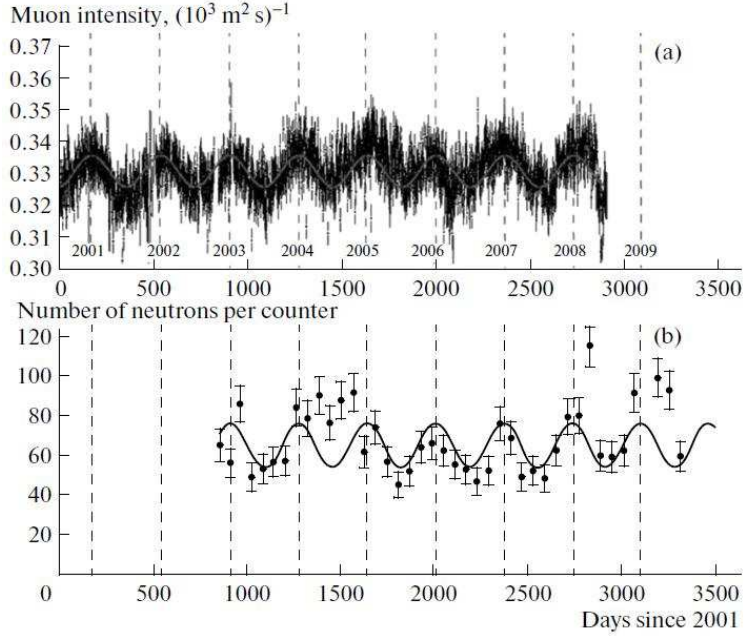}
  \caption{(a) Muon intensity variations per day over 8 years of LVD
           operation. (b) The number of neutrons from muons per counter;
           each point represents the data obtained over two months of 
           LVD operation.}
  \label{1fig}
 \end{figure}

The variations in the number of neutrons generated by the muon flux 
in the detector material were detected at the LVD \cite{Aga11}. 
The parameters of variations were determined using the LVD data 
from April 1, 2003, to April 1, 2010. To increase the 
statistics, the number of muon-induced neutrons was determined with a 
step of 60 days (Fig. 1b). When fitting the experimental data by a 
function
\begin{equation}
N(t) = N_0 + {\delta}N cos(\frac{2{\pi}}{T}(t-t^n_0))
\end{equation}
the best agreement of the fitting curve with the experimental 
data points is achieved at the following parameters of the 
function: $N_0 = 65.0 \pm 2.2$, ${\delta}N = 9.3 \pm 3.9$, $T = 1$ yr, 
and $t^n_0= 185 \pm 18$ days. The derived phase agrees with the phase 
of muon variations $t^{\mu}_0$ \cite{Aga11}.

The maximum relative increase in the number of neutrons is
\begin{equation}
k_n = \frac{N^{max}}{N_0} = 1 + \frac{{\delta}N}{N_0} = 1.143.
\end{equation}
A similar value can be obtained for the muon intensity using the 
parameters of function (1):
\begin{center}
$k^I_n = 1.015$.
\end{center}
It has been pointed out above that the number of neutrons detected 
by the LVD in 60 days must depend both on the number of muons passed 
through the detector in this time and on their energy. In such a case,
\begin{equation}
k_n = k^I_nk^E_n,
\end{equation}
whence
\begin{equation}
k^E_n =\frac{k_n}{k^I_n} = 1.126,
\end{equation}
$k^E_n$ is a coefficient that allows for the maximum change in 
energy $\overline E_{\mu}$. It is well known that the dependence of 
the number of neutrons on muon energy can be described by a 
power law $E^{\alpha}_{\mu}$, consequently,
\begin{equation}
k^E_n =(\frac{E^{max}_{\mu}}{\overline E_{\mu}})^{\alpha}.
\end{equation}
Hence we determine the maximum mean energy of the muon flux 
(summer value) as a function of the yearly mean $\overline E_{\mu}$:
\begin{equation}
E^{max}_{\mu} = (k^E_n)^{1/{\alpha}}\overline E_{\mu}.
\end{equation}
The quantity ${\alpha}$ has been investigated theoretically and 
experimentally; it is limited by 0.7 and 0.8 \cite{Rya65}, \cite{Wan01}, 
\cite{Ara05}. The best agreement with the experimental data on the 
neutron yield is observed at ${\alpha}$ = 0.78 \cite{Aga13}.
Substituting this value into (7), we find at 
$\overline E_{\mu}$ = 280 GeV that 
$E^{max}_{\mu}$ = 326 GeV.
Thus, assuming that the seasonal variations of the muon energy and 
intensity have the same origin and that the deviations 
$E^{max}_{\mu} - \overline E_{\mu}$  and $\overline E_{\mu} - E^{min}_{\mu}$
 are equal (because ${\alpha}$ is close to 1), we obtain the time 
dependence of the energy $E_{\mu}(t)$ averaged over the muon flux in 
general form:
\begin{equation}
E_{\mu}(t) = \overline E_{\mu} + {\delta}E_{\mu}cos[\frac{2{\pi}}{T}(t-t^{\mu}_0)].
\end{equation}
At the LVD depth, $\overline E_{\mu} = 280$ GeV and ${\delta}E_{\mu}$ = 46 GeV; 
the relative modulation amplitude of the energy averaged over the muon flux is 16\%.
The derived amplitude ${\delta}E_{\mu}$ = 46 GeV has an uncertainty of 
$\sim 60\%$, which is attributable mainly to the error in ${\delta}N$, $\sim 40\%$.

Being dependent on the intensity $I^{\mu}$ and energy $E_{\mu}$, the neutron 
flux also undergoes seasonal variations. The annual mean neutron flux at a 
given depth $H$ is expressed by the formula
\begin{equation}
F_0(H) [n \cdot cm^{-2} s^{-1}] = I^{\mu}_0(H)Y(\overline E_{\mu}){\lambda}_n,
\end{equation}
$I^{\mu}_0(H) [\mu \cdot cm^{-2} s^{-1}]$ is the annual mean global muon intensity 
at depth $H$; $Y(\overline E_{\mu}, A) [n / {\mu} /(g \cdot cm^{-2})]$
is the cosmogenic neutron yield in a material with mass number $A$ at muon energy [GeV] 
corresponding to this depth; ${\lambda}_n [g \cdot cm^{-2}]$ is the cosmogenic neutron 
attenuation length ($\sim 40 g \cdot cm^{-2}$ for standard rock $A$ = 22, $Z$ = 11).
Using the formula for the neutron yield obtained in \cite{Aga13},
\begin{equation}
Y(\overline E_{\mu}, A) = b \overline E^{0.78}_{\mu}A^{0.95}, b = 4.4 \times 10^{-7} cm^2 g^{-1},
\end{equation}
we arrive at an expression for the neutron flux in a material 
with mass number $A$:
\begin{equation}
F_0(H) = b{\lambda}_nI^{\mu}_0(H)\overline E^{0.78}_{\mu}A^{0.95}.
\end{equation}
Taking into account $k_n = 1.143$, we obtain an expression 
for the seasonal modulations in the flux of cosmogenic neutrons at 
the LVD depth $H_0$ produced in material $A$:
\begin{equation}
F_0(t) = F_0(H_0) [1+0.143cos(\frac{2{\pi}}{T}(t-t^n_0))]. 
\end{equation}
An enhancement of the annual modulations in the cosmogenic neutron flux 
compared to the muon flux arouses a desire to associate the signal modulations 
in the DAMA/LIBRA experiment \cite{Ber13} with them. The difference in the 
modulation phases of the neutron flux $t^n_0 = 185 \pm 18$ days (the maximum occurs 
at the beginning of July) and the DAMA/LIBRA signal $t^{D/L}_0 = 152.5$ days 
(the maximum occurs on June 2) contradicts this. 

\section{Conclusions}

The variations ${\delta}E_{\mu}$ in the mean energy of the muon flux 
are the main source of seasonal variations in the underground cosmogenic 
neutron flux. The relative amplitude of the neutron variations related to
${\delta}E_{\mu}$ exceeds the relative modulation amplitude ${\delta}I_{\mu}$
by an order of magnitude. This is explained by a stronger dependence of the 
energy on the temperature effect than that for $I_{\mu}$.

Taking into account the dependence of the temperature coefficient ${\alpha}_T$
on depth (the decrease of ${\alpha}_T$ with decreasing depth), which relates 
the seasonal muon intensity variations ${\Delta}I^{\mu}/I^{\mu}$ 
to the atmospheric temperature variations  ${\Delta}T/T$ \cite{Dav14}, \cite{Bar52},
\cite{Amb97}, one should expect a corresponding depth dependence of the coefficient 
$k^E_n$ and, as a consequence, $E^{max}_{\mu}$.

Apart from the seasonal modulations, the temperature of the upper atmospheric layers 
undergoes irregular variations over a year. As a result, the number of neutrons 
produced by muons underground deviates considerably from the harmonic function 
(2) with a breakdown of the constancy of the modulation amplitude ${\delta}N$ and
the fluctuation phase $t^n_0$. This should be taken into account when analyzing the 
background in low-background underground experiments.

\end{document}